\documentclass[pdftex]{bioinfo}
\copyrightyear{2014}
\pubyear{2014}
\usepackage{hyperref}
\usepackage[numbers]{natbib}
\begin{document}
\firstpage{1}

\title[ViDaExpert: user-friendly tool for nonlinear visualization and analysis of multidimensional vectorial data]{ViDaExpert: user-friendly tool for nonlinear visualization and analysis of multidimensional vectorial data}
\author[Gorban \textit{et~al}]{Alexander N. Gorban$^{1,2}$, Alexander Pitenko$^{2}$ and Andrei Zinovyev$^{3,4,5}$ \footnote{to whom correspondence should be addressed}}
\address{
$^{1}$University of Leicester, UK\\
$^{2}$Institute of Computational Modeling, SB RAS, Krasnoyarsk, Russia\\
$^{3}$Institut Curie, Paris 75248 France\\
$^{4}$INSERM U900, Paris 75248 France\\
$^{5}$Mines Paris Tech, Fontainebleau, 77300 France}

\history{Received on XXXXX; revised on XXXXX; accepted on XXXXX}

\editor{Associate Editor: XXXXXXX}

\maketitle

\begin{abstract}
ViDaExpert is a tool for visualization and analysis of multidimensional vectorial data. ViDaExpert is able to work with data tables of "object-feature" type that might contain numerical feature values as well as textual labels for rows (objects) and columns (features). ViDaExpert implements several statistical methods such as standard and weighted Principal Component Analysis (PCA) and the method of elastic maps (non-linear version of PCA), Linear Discriminant Analysis (LDA), multilinear regression, K-Means clustering, a variant of decision tree construction algorithm. Equipped with several user-friendly dialogs for configuring data point representations (size, shape, color) and fast 3D viewer, ViDaExpert is a handy tool allowing to construct an interactive 3D-scene representing a table of data in multidimensional space and perform its quick and insightfull statistical analysis, from basic to advanced methods.

\section{Availability:}
ViDaExpert software is freely available at
\url{http://bioinfo-out.curie.fr/projects/vidaexpert}. The tool does not require
installation. Currently, there is no implementation of ViDaExpert for platforms other
than Windows (any version).

\section{Contact:} \href{zinovyev@gmail.com}{zinovyev@gmail.com}

\section{Supplementary Information:}
1) Tutorial slides representing the major functions of ViDaExpert

\end{abstract}

\section{Introduction}

ViDaExpert (ViDa stands for Visualization of Data) is a software implementing a number of
simple and advanced statistical methods and a user-friendly graphical user interface
(GUI) for applying these methods to a table of data which can contain both numerical
feature values and labels for objects and features. The primary objective of ViDaExpert
is to implement a user interface to the method of {\bf elastic maps} for non-linear data
dimension reduction and visualization developed by the authors of this paper
\cite{Gorban2000_Mashinostroenie, SESD2001, CHAOS2001, VisBook2000,
Zinovyev_NeuroComputeri2002, Gorban_NeuroComputeri2002, Computing2005, AML2007, Book2008,
HandBook2009, IJNS2010}. It appeared advantageous to introduce standard methods of
multivariate statistical analysis into the software to be able to visualize the result of
their applications on the projections onto the elastic map (non-linear principal
manifold). Currently, ViDaExpert contains quite a diverse set of statistical tools
making it a handy self-consistent environment for performing fast exploratory statistical
analysis and visualization of multivariate data.

\begin{figure*}[screenshot_elmap]
\begin{center}
\includegraphics[width=0.9\textwidth]{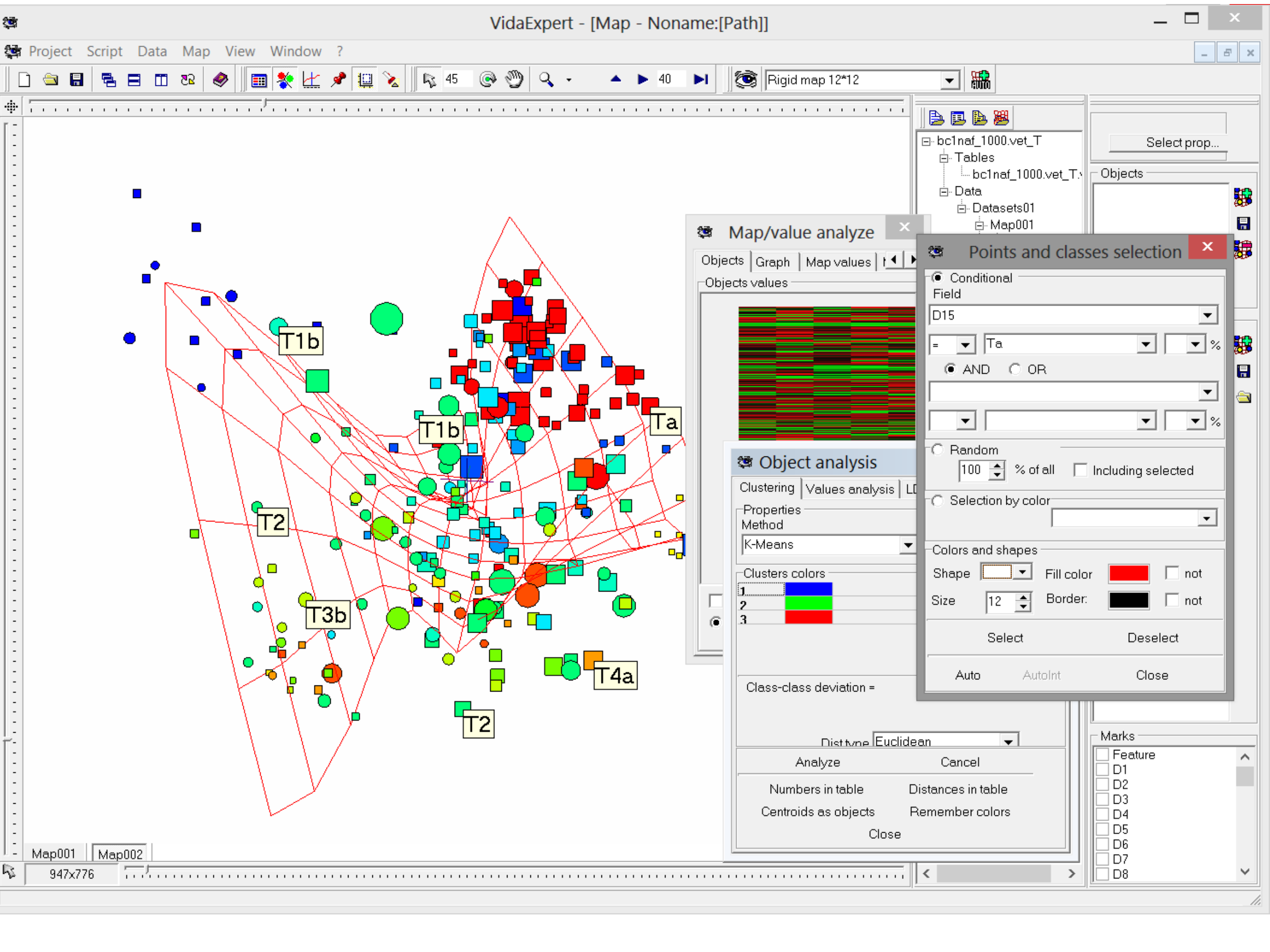}
\caption{ViDaExpert GUI screenshot. Most of the working space is occupied by interactive
3D-scene visualizing a cloud of data points and the elastic maps constructed for these
data. Several specialized dialogs allows to launch various statistical procedures and
visualize the results of their applications in the scene.} \label{SnapShot_elmap}
\end{center}
\end{figure*}

\begin{figure*}[screenshot_pcurve]
\begin{center}
\includegraphics[width=178mm]{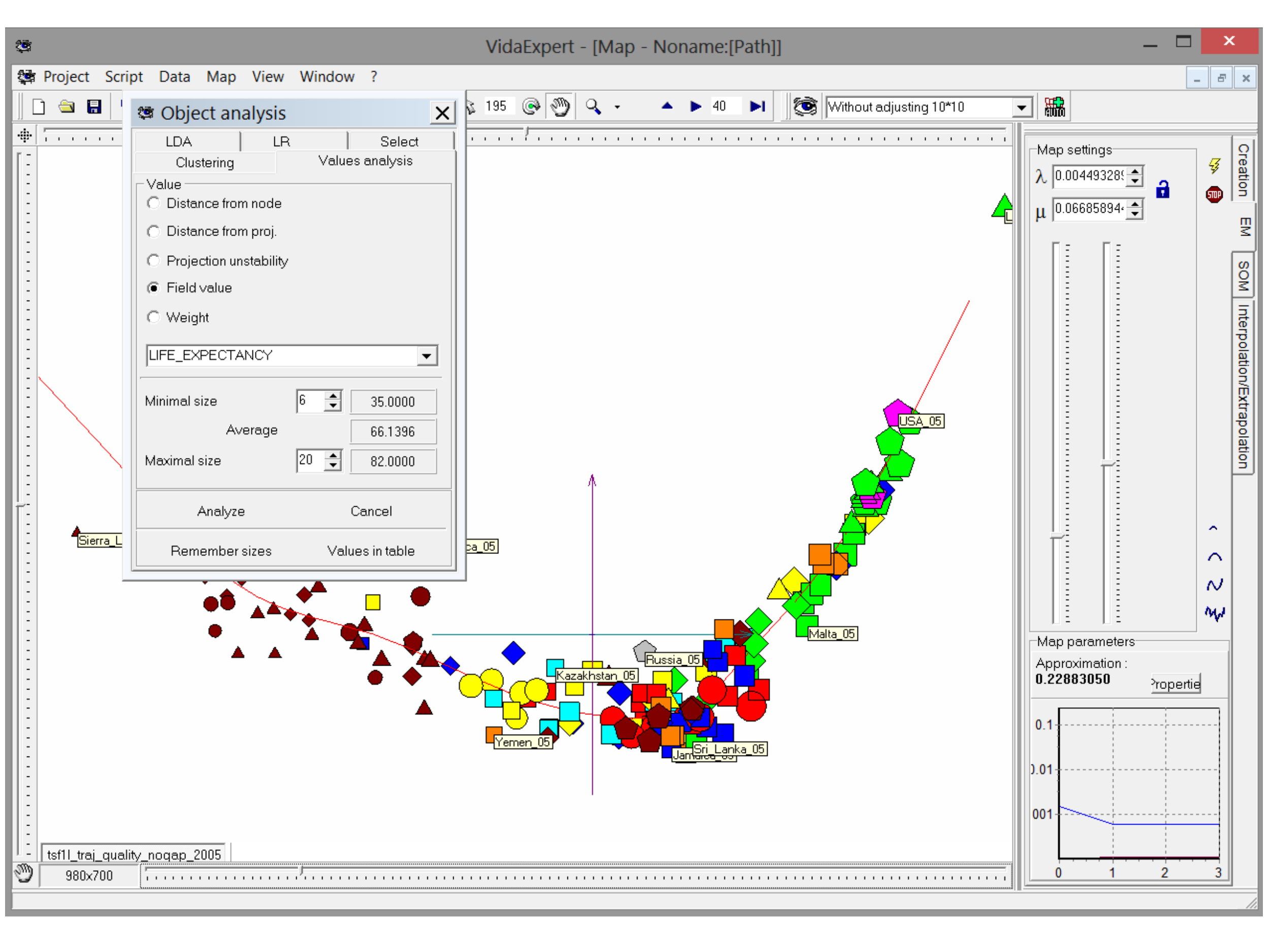}
\caption{Using VidaExpert for constructing a principal curve for a dataset of quality of life indices for all countries in 2005 \cite{IJNS2010,Index2010}.}
\label{SnapShot_pcurve}
\end{center}
\end{figure*}

\begin{figure*}[screenshot_biplot]
\begin{center}
\includegraphics[width=178mm]{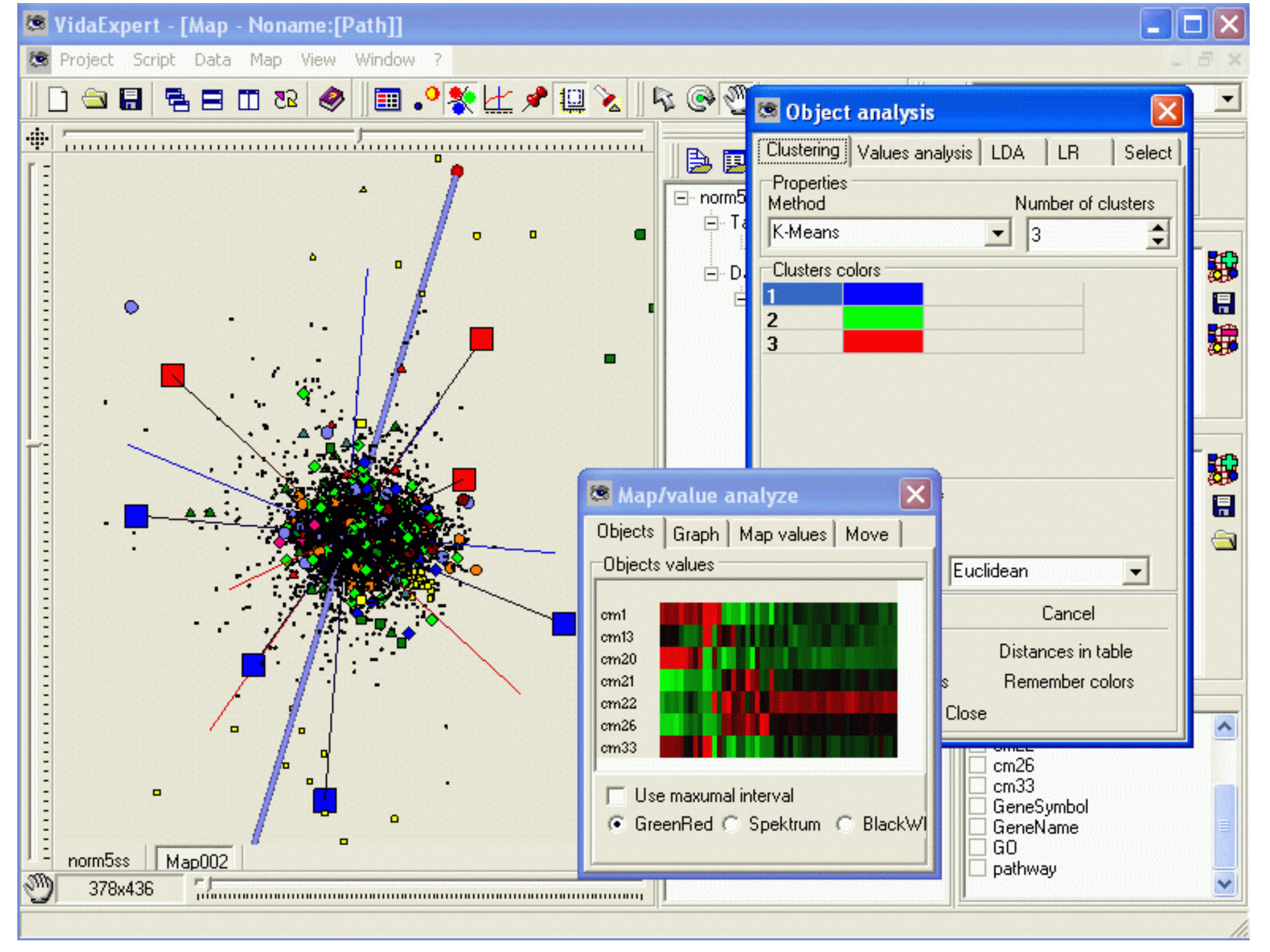}
\caption{Constructing biplot representation of principal components \cite{Gabriel1971} in VidaExpert for a distribution of gene expression values.}
\label{SnapShot_biplot}
\end{center}
\end{figure*}

\begin{figure*}[screenshot_coloring]
\begin{center}
\includegraphics[width=178mm]{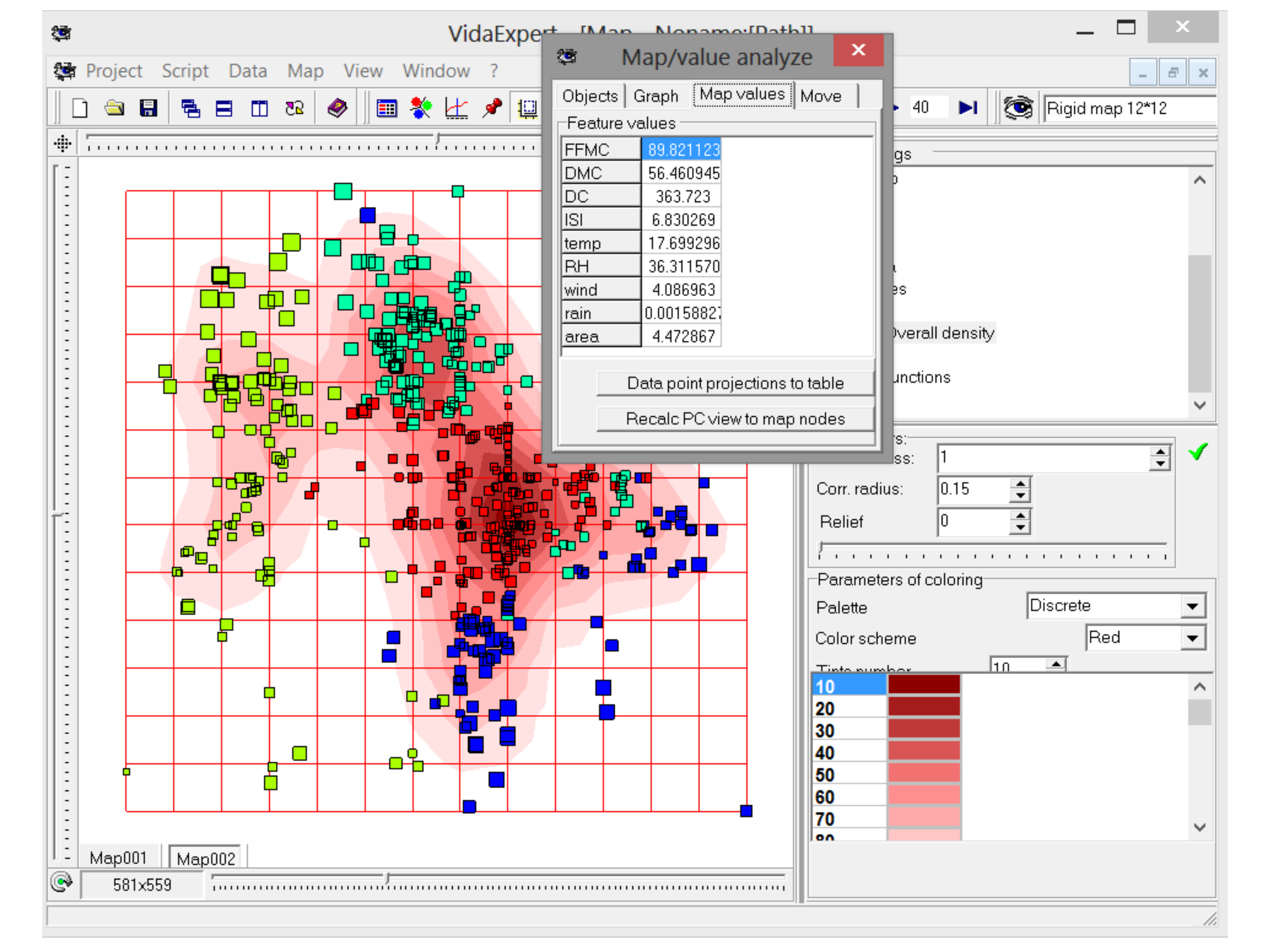}
\caption{Visulializing the density of data point projections onto the principal manifold (gradient in the background) in order to
allow the visual estimation of existence of subgroups in the dataset describing forest fires \cite{Cortez2007} and
comparing it with the results of K-Means clustering algorithm (shown by coloring the data points). }
\label{SnapShot_coloring}
\end{center}
\end{figure*}

\begin{figure*}[screenshot_molsurf]
\begin{center}
\includegraphics[width=178mm]{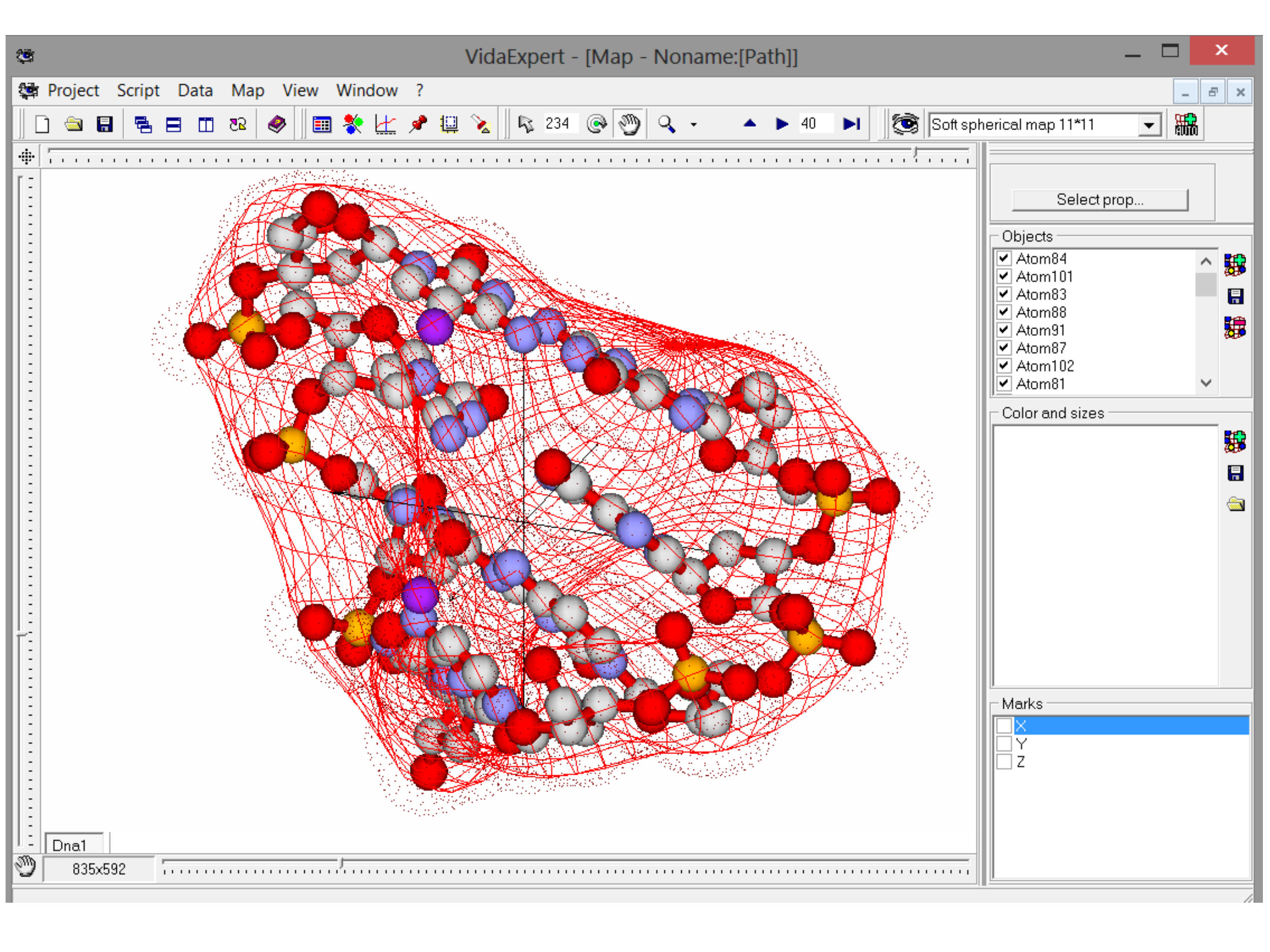}
\caption{Using VidaExpert for visualizing the backbone of a molecule (several DNA nucleotides in this case) in 3D,
and approximating the continuous molecular surface by computing the principal manifold with spherical topology \cite{Computing2005}.}
\label{SnapShot_molsurf}
\end{center}
\end{figure*}

The core of the ViDaExpert GUI is a fast viewer allowing to smoothly interact with a
3D-scene representing a cloud of points (objects) in multidimensional space of table
feature values (Figure~\ref{SnapShot_elmap}). The scene can include points themselves
together with other objects that can be introduced into the scene: principal manifolds,
vectorial fields, structure of biological molecules with spheres and cylinders, etc. The
appearance of objects can be easily configured to represent various numerical values and
object labels. A user can interact with the scene not only by rotating it but also
selecting data points and defining specific actions for them (for example, assigning a
new color).

ViDaExpert and method of elastic maps are intensively used in bioinformatics, sociology,
political sciences and many other domains. A brief review of known applications of
ViDaExpert is provided at the end of this paper.

\section{Basic principles of ViDaExpert software}

The basic scenario of working in VidaExperts consists of several simple steps.

(1) The user data should be prepared in a specific text-based format described below and
loaded into ViDaExpert GUI. This will create an object called further DataTable.

(2) From DataTable, a user should create a DataSet: an object representing the cloud of
points in multidimensional data.

(3) DataSet can be visualized using one or several Map objects representing a mapping of a multidimensional data into a low-dimensional space.

(4) A user can apply various statistical methods implemented in ViDaExpert, such as Linear Discriminant Analysis, and see visualization of the results of these methods in the constructed 3D-scene.

The steps of this scenario are explained and illustrated in the Tutorial in Supplementary Materials for this text.

\subsection{Input format}
The recommended format for using in ViDaExpert is a textual file having ".dat" extension.
The file represents the table of numerical and string values with a simple header which
contains, first, type of the column (FLOAT for numerical type and STRING for the labeling
information) and, second, any labeling of columns (see example in the tutorial). Importantly, it is
recommended to put in quotes any textual information and use tab symbol to delimitate the
columns in the table and in the header.

\subsection{DataTable object}

When .dat file is loaded into ViDaExpert, a DataTable object is created. In one
ViDaExpert session, a user can load several DataTable objects and switch between them. A
user can save the content of the table into ViDaExpert format, which will have ".vet"
extension. In this case, all attributes associated to objects (color, shape and size)
will be saved as well and can be restored later.

\subsection{DataSet object}

DataSet represents a numerical matrix in ViDaExpert. This matrix is formed by selecting certain columns having numerical values in DataTable and normalizing (scaling) them. The default normalization consist of subtracting the mean value and dividing by standard deviation of the values in the column. Other types of normalization are also available, including reducing the values to [-1;1] interval, using logarithm or hyperbolic tangent function. Several types of normalization can be combined for creating a DataSet. In addition, the user can decide not to use those rows in DataTable which contains undefined numerical values (gaps): in this case the number of rows in DataSet will become smaller than the number of rows in DataTable. However, many methods in ViDaExpert (including PCA and elastic maps) are capable of dealing with moderate number of gaps in a DataSet, even without prior imputation of missing values.

For each DataTable, a user can create several DataSets, which might be advantageous for using the same set of object attributes (color, size and shape) for showing the data points in different data spaces created by different column selections. For example, colors defined from application of K-Means clustering in one data space can be visualized in another data space, which was not used for clustering.

The DataSet can be saved into the ViDaExpert format having ".ved" extension and loaded later.

\subsection{Map object}

The Map object is a representation of an elastic map, low-dimensional manifold (screen), computed for a cloud of multidimensional vectors represented by a DataSet object. By default, the map represents a linear manifold spanned by the direction of the first two principal components. A user can change a number of predefined parameter configurations in order to compute a principal manifold by elastic map method, or construct the elastic map in a manual mode, by specifying the elastic parameters of the map.

After the map is created, a user can visualize both data points and the map together in several projections: (1) in the space of the three chosen by the user data space coordinate axes; (2) in the space defined by first three principal components and (3) in the internal space of the map after projecting the data points onto it. In all cases, the constructed 3D-scene can be rotated, zoomed and shifted by the user.

The map can be saved into the ViDaExpert format having ".vem" extension and loaded later.

\section{Using Principal Component Analysis in ViDaExpert}

Principal Component Analysis (PCA) is used in ViDaExpert for several purposes: (1) for constructing one of the projections of the data points and the elastic maps; (2) for initializing the elastic map; (3) for analysing contribution of a feature into the $n$th principal component and estimating the amount of variance explained by the $n$th principal component. Implementation of PCA in ViDaExpert differs from most of the standard implementations in the following aspects: (1) it is able to work with weighted data vectors (there is a dialog in ViDaExpert allowing to associate a weight to each data point) and (2) it is able to compute principal components for a matrix which can contain missing values without imputing them. The iterative algorithm for computing Singular Value Decomposition (SVD) for PCA is described in details in \cite{HandBook2009}.

For user convenience, ViDaExpert allows to make standard DataTable transformations, frequently utilized in PCA such as transposing the DataTable or subtracting the first principal component (which can help to eliminate, for example, a major bias affecting the average feature values of an object, through all columns). Another useful implemented feature is using biplots for presenting the results of linear PCA \cite{Gabriel1971} (see Figure~\ref{SnapShot_biplot}).

\section{Elastic maps method and its use in ViDaExpert}

Elastic map provides a method for nonlinear dimensionality reduction. Elastic map is a
system of elastic springs embedded in the data space and approximating a low-dimensional
manifold. Tuning the elastic coefficients of springs allow switching from completely
unstructured k-means clustering (zero elasticity) to the estimators located closely to
linear PCA manifolds (for high bending and low stretching elasticities). With some
intermediate values of the elasticity coefficients, this system effectively approximates
non-linear principal manifolds. The approach is based on a mechanical analogy between
principal manifolds, that are passing through "the middle" of the data distribution, and
the elastic membranes and plates. The method was developed by the authors of this paper,
starting from 1996 -– 1998. The most exhaustive theoretical description of the elastic map
methodology is provided in \cite{HandBook2009} together with a comprehensive review on
the related methods. A more practice-oriented description is given in \cite{IJNS2010}.

\subsection{Constructing the principal manifold}

Construction of the elastic map is based on expectation-minimization-based energy
optimisation algorithm, which utilises annealing methodology in order to achieve a deeper
local minimum of the energy function. Therefore, the elastic map is trained in several
epochs each characterized by certain elasticity coefficients. After optimising the
position of grid nodes of the system of springs, the manifold is extended by linear
extrapolation to its vicinity in the data space, in order to avoid projecting many data
points onto the border of the manifold. In ViDaExpert, the sequence of these steps is
either predefined in a number of standard scenarios which can be applied by a single
click, or can be set manually by a user.

Elastic map algorithm allows constructing principal manifolds of various topology
(rectangular, hexagonal, spherical) and dimension (1D manifolds or principal curves, 2D
or 3D manifolds). Several such possibilities are provided to the ViDaExpert users.
Among these build-in scenarios there are:

"Without adjusting" computes simple 2D linear principal manifold.

"Rigid map" computes relatively rigid and smooth elastic map, closer to the linear one.

"Soft map" computes much more non-linear than "Rigid map" elastic map, which better
approximates the data, but can be trapped in some too complicated locally optimal
configurations.

"Soft spherical map" constructs a 2D principal manifold having topology of a sphere.

"Detailed map" constructs a systems of springs as a grid with many nodes ($25\times
25=625$ nodes) which is bigger than typically used in other scenarios ($10\times 10 =100$
nodes).

"3D" map constructs three-dimensional non-linear principal manifolds by using a grid with
$6\times 6\times 6=216$ nodes.

\subsection{Coloring the manifold}

After the manifold is constructed, it can be used in a number of ways. First of all, it
can be used to visualize the distribution of data points after projecting them into the
closest point of the manifold, and visually estimate if the points form a cluster
structure (which in practice can be fuzzy and can not be easily determined by clustering
algorithms). Second, the manifold itself can be used to visualize some functions defined
in the multidimensional space or in the space of internal manifold coordinates.

In this way, we implemented a possibility to visualize (1) local density of data points
both in multidimensional and in the projected space; (2) the feature values in each
point of the manifold that represents a smoothed trend of the feature along the
manifold; (3) some other functions such as those resulting from application of linear
regression (LR) or linear discriminant analysis (LDA).

This technique is called "Map coloring" in ViDaExpert.
For example, coloring by density allows to visually perform clustering of data points (Figure~\ref{SnapShot_coloring}).
Coloring by a particular feature value produces a smoothed trend image of the values of this feature, along the manifold.
Coloring by the result of LDA allows to visualize the distribution of misclassified data
points. The user can use discrete or gradient coloring, choose tints of several
pre-defined colors, use spectral-type coloring in which blue color
denotes smaller and red color denotes bigger values, and use green-red coloring which is
typically used in bioinformatics applications.

\subsection{Interacting with the manifold}

In ViDaExpert, a user can interact with constructed elastic manifold, using specialized
dialogs. For example, it is possible to shift (translate) or rotate the manifold along
one of the linear axes defining the three-dimensional subspace used for visualization. It
is also possible to compute interactively the projection onto the manifold of a point in
the dataspace defined by mouse pointer (this answers the question "what would be
projection of a data point if it appears in the position of the mouse pointer").

\section{Usefull ViDaExpert features}

\subsection{Configuring appearance of objects}

ViDaExpert contains advanced dialogs containing multiple possibilities to change
appearance of the objects in order to reflect their labels or numerical values by using shape,
color (both background and the border colors of the points) and size. Many of these
possibilities are automated: for example, it is possible to assign, in one click,
different colors to data points for each distinct text they have in a certain label
(point class information). Two features of the DataTable can be combined by OR or AND
logics in order to define a certain visual appearance of data points.

In most scenario, the background color of data points is associated to the class or
cluster information. In contrast, the size of the points can be used to map the values of
certain features, or some other values such as the distance to the constructed manifold
(approximation residue).

\subsection{Labeling objects}

One of the user dialogs in ViDaExpert allows the user to attach a label to all or to a
subselection of objects (data points). This label can reflect one of the row labels
contained in the DataTable, a numerical value of a feature, or combine several labels
together. In addition, more advanced labels can be assigned to data points, such as the
name of the "least expected feature" (the feature that has the
least typical value for this point).

\subsection{Interaction with Microsoft Excel software}

A usefull feature of ViDaExpert is using OLE technology to call Excel software and
communicate some table information without saving it on the disk. Most of the dialogs for
the implemented methods in ViDaExpert are equipped with "In Excel" button, allowing to
transfer the results of the method's application (i.e., the feature contributions into
the first three principal components) into Excel, where the user can create different
plots or manipulate the data at his discretion.

\subsection{Introducing auxiliary objects into ViDaExpert 3D-scene}

ViDaExpert allows describing a scene composed from spheres and cylinders positioned in
the multidimensional space and then projected into a low-dimensional space. The
corresponding file describing positions, colors and sizes of objects has ".veo" extension ("o" stands for "objects"). This possibility
might be usefull in many situations, for example, for marking some distinguished points
(centroid of a class) in the space and connections between them. Another application is
visualizing the structures of complex molecules in ViDaExpert, using its 3D-viewer for
rotating them (Figure~\ref{SnapShot_molsurf}). Another possibility consists in introducing line segments into the
ViDaExpert scene, which, for example, can be used for visualization of a vector field.

\section{Statistical methods implemented in ViDaExpert}

Application of statistical methods in ViDaExpert is accompanied by visualization of the
results in the 3D scene of ViDaExpert. Usually, the data points are changing their
appearances in the scene, reflecting application of a method. These appearances are
temporary: appearance of data points before application of the method is memorized and
can be restored by clicking the "Cancel" button in any dialog devoted to a statistical
method. However, the temporary appearance can be fixed into the permanent one by clicking
"Remember colors" or "Remember sizes". In addition, the dialogs allow to write the
results into DataTable, by creating a separate column (for example, containing the
cluster number) or export the results to Excel software.

\subsection{Clustering methods}

ViDaExpert contains the most basic implementation of K-Means algorithm. The clusters
found by K-Means are assigned different colors such that the user can fix them or
investigate them in the 3D scene. There are alternative ways of clustering possible,
for example, by assigning a point to the closes node of the elastic map (analogously
to Self-Organizing Maps clustering).

\subsection{Linear Discriminant Analysis}

In ViDaExpert one of the simplest and classical versions of LDA is implemented. Before
application of the method, the user should define the color of the data points which he wants to
separate from the rest of the data cloud. After constructing the separation function,
ViDaExpert reports on the achieved sensitivity and specificity of classification and
visualize the distribution of misclassified data points by changing their color and size,
accordingly to the distance to the separating hyperplane.

\subsection{Multiple Linear Regression Analysis}

In order to perform linear regression, the user specifies which feature should be fitted
by linear combination of other features and the allowed error interval. When the regression function is computed,
ViDaExpert reports on the number of points whose regression residues are not bigger that
the specified interval (the quality of regression). These results are immediately
reflected in the 3D-scene of ViDaExpert.

\section{Other usefull features}

Among other frequently used features of ViDaExpert, one can mention the following ones.

A user can construct histograms of table column values; the plots of the histograms are
zoomable and clickable such that a user can mark data points belonging to a certain
histogram bar.

A user can investigate distance distribution from one object to all other objects, which
can help in answering the question if the position of the object is atypical (like an outlier) or
it is located in the middle of dense cloud of other points. One can distinguish points of
different classes (colors) in this study.

There is a possibility to construct a bi-plot representation of the PCA analysis (Figure~\ref{SnapShot_biplot}).

Several original features of ViDaExpert were developed for presenting the result
ViDaExpert application to data in scientific presentations. For example, there is a possibility to create an
animated GIF image from a series of bitmaps representing a rotating 3D scene from ViDaExpert.

\section{Applications}

ViDaExpert software has been applied in many domains of science where there is a need to
visually represent tables of numerical data.

Thus, it was used to visualize economic indicators of Russian economy
\cite{Gorban2000_Mashinostroenie, Zinovyev_NeuroComputeri2002, SESD2001}. ViDaExpert was
applied in political science for data visualization \cite{Encyc2011, IJNS2010}. In
particular, this technology solves a classical problem of unsupervised ranking of
objects. It allows to find the optimal and independent on expert's opinion way to map
several numerical indicators from a multidimensional space onto the one-dimensional space
of the ``quality'' or ``index'' \cite{Index2010,IndexChina2014} (see Figure~\ref{SnapShot_pcurve}).

The method is adapted as a support tool in the decision process underlying the selection,
optimization, and management of financial portfolios \cite{Resta2010}.

Most of the applications of ViDaExpert software and elastic map method found in
bioinfomatics. It was used to visualize the universal 7-cluster structure of bacterial
genomes \cite{IJCNN2003,ISB2003} and the structure of codon usage in genomes of various
organisms \cite{Carbone2003,ZinovyevHDR2014}. Elastic maps allow approximation of molecular surfaces of
complex molecules and visualizing them in ViDaExpert \cite{Computing2005} (see Figure~\ref{SnapShot_molsurf}). ViDaExpert is
routinely used for analysis of microarray data in cancer biology
\cite{ChapterElMap2008,IJNS2010,ZinovyevHDR2014} and in biology of microorganisms \cite{Chacon2007}. The
method of elastic maps is applied in quantitative biology for reconstructing the curved surface of a tree
leaf from a stack of light miscroscopy images \cite{Failmezger2013}.

The method of elastic maps was successfull in tracing skeletons of handwritten symbols
\cite{Computing2005}. The method of elastic maps has been systematically tested and
compared with several machine learning methods on the applied problem of identification
of the flow regime of a gas-liquid flow in a pipe \cite{Shaben2014}. Generalizations of
elastic map method can be used to quantify and compare the complexity of large sets of data
\cite{ZinovyevMirkes2013,IWANN2013,ZinovyevHDR2014}.

\section{Future development}

ViDaExpert will be further developed in order to implement
methods of non-linear dimension reduction and visualization which stemmed from the
elastic map method (such method of principal trees, metro map data visualization),
described in more details in \cite{IJCNN2007, AML2007, ChapterPrincTree2008,
HandBook2009}. Another direction is implementation of other methods of linear factor
analysis such as Independent Component Analysis \cite{Zinovyev2013_BBRC}.

\section{Implementation}

ViDaExpert code is written in Delphi language (Object Pascal). For 3D visualization,
OpenGL Windows library is used.

\end{document}